\documentclass[aps,prc,reprint,nofootinbib,superscriptaddress,reprint,floatfix]{revtex4-2}
\usepackage{graphicx}
\usepackage{dcolumn}
\usepackage{bm}
\usepackage{amssymb}
\usepackage{amsmath}
\usepackage{amsfonts}
\usepackage{xcolor}
\usepackage{multirow}
\usepackage{float}
\usepackage{tabularx}
\usepackage{amsbsy}
\usepackage{graphicx}
\usepackage{hyperref} 
\usepackage{lmodern}
\usepackage[T1]{fontenc}
\usepackage{bigints}
\usepackage{comment}

\def\empile#1\over#2{\mathrel{\mathop{\kern 0pt#1}\limits_{#2}}}

\def\beq{\begin{equation}}
\def\eeq{\end{equation}}
\def\bea{\begin{eqnarray}}
\def\eea{\end{eqnarray}}

\newcommand{\Lb}{\left(}
\newcommand{\Rb}{\right)}

\begin{document}


\title{\Large Understanding the systematic differences in extractions 
of the proton electric form factors at low-\texorpdfstring{$Q^2$}{Q2}}



\author{Jingyi Zhou}
\affiliation{Department of Physics, Duke University, Durham, North Carolina 27708, USA}
\affiliation{Triangle Universities Nuclear Laboratory, Durham, North Carolina 27708, USA}

\author{Vladimir Khachatryan}
\email{vladimir.khachatryan@duke.edu}
\affiliation{Department of Physics, Duke University, Durham, North Carolina 27708, USA}
\affiliation{Triangle Universities Nuclear Laboratory, Durham, North Carolina 27708, USA}

\author{Haiyan Gao}
\affiliation{Department of Physics, Duke University, Durham, North Carolina 27708, USA}
\affiliation{Triangle Universities Nuclear Laboratory, Durham, North Carolina 27708, USA}

\author{Simon Gorbaty}
\affiliation{Department of Physics, Duke University, Durham, North Carolina 27708, USA}
\affiliation{Triangle Universities Nuclear Laboratory, Durham, North Carolina 27708, USA}

\author{Douglas W. Higinbotham}
\affiliation{Thomas Jefferson National Accelerator Facility, Newport News, Virginia 23606, USA}

\begin{abstract}
Systematic differences exist between values of the proton's electric form factors in the low-$Q^2$ region extracted by different experimental 
and theoretical groups, though they are all making use of basically the same electron-proton scattering data. To try understand the source of 
these differences, we make use of the analytically well-behaved rational (N=1, M=1) function, a predictive function that can be reasonably used 
for extrapolations at $Q^{2} \rightarrow 0$. First, we test how well this deceptively simple two-parameter function describes the extremely 
complex and state-of-the-art dispersively improved chiral effective field theory calculations. Second, we carry out a complete re-analysis of 
the 34 sets of eletron-proton elastic scattering cross-section data  of the Mainz A1 Collaboration with its unconstrained 31 normalization 
parameters up to $Q^{2} = 0.5~{\rm (GeV/c)^{2}}$. We find that subtle shifts in the normalization parameters can result in relatively large 
changes in the extracted physical qualities. In conclusion, we show that by simply using a well-behaved analytic function, the apparent
discrepancy between recent form-factor extractions can be resolved.
\end{abstract}

\maketitle



\section{\label{sec:intro} Introduction}

Nucleons are the building blocks of atomic nuclei, which constitute essentially all visible matter in the universe. Therefore,  
understanding their composition and dynamics in terms of the underlying quark-gluon degrees of freedom of Quantum Chromodynamics (QCD) -- 
the theory of strong interaction -- has been at the frontier of modern nuclear and hadronic physics for decades. The electromagnetic 
structure of the nucleon is traditionally considered to be directly accessible by the Sachs electric ($G_{E}^{p}$) and magnetic 
($G_{M}^{p}$) form factors in the proton case. Form-factor studies have attracted tremendous interest for more than half a century, as 
demonstrated by enormous experimental and theoretical efforts since the 1950s and 1960s 
\cite{Hofstadter:1956qs,Ernst:1960zza,Sachs:1962zzc,Hand:1963zz}
to recent review articles~\cite{Punjabi:2015bba}.
The precise information about nucleon electromagnetic form factors allows for precision tests of lattice QCD calculations. The proton 
charge radius extracted from the proton electric form factor is an important input to the bound state Quantum Electrodynamics (QED) 
calculations of atomic energy levels and to the determination of the Rydberg constant, one of the most precise fundamental constants 
in physics \cite{Tiesinga:2021myr}. The recent substantial progress in our understanding of the form factors, extracted from electron-proton 
elastic scattering can be found in \cite{Arrington:2006zm,Perdrisat:2006hj,Punjabi:2015bba, Gao:2021sml}. 

The commonly applied technique to extract $G_{E/M}^{p}$ from an unpolarized $e-p$ elastic scattering cross section is the
Rosenbluth separation method~\cite{Rosenbluth:1950yq}. This was the case for most of and especially early proton 
form factor measurements. This method works well in a kinematic region, where both $G_{E/M}^{p}$ contribute to the cross section significantly, 
and as such, these form factors can be extracted without introducing large systematic uncertainties associated the method itself. However, 
the $G_{E}^{p\,2}$ term will dominate the cross section if $Q^{2}$ is small, whereas the $G_{M}^{p\,2}$ term becomes dominant at high $Q^{2}$. 
As a result, in a high-$Q^{2}$ region the $G_{E}^{p}$ data typically have large uncertainties, while the $G_{M}^{p}$ data uncertainties 
are larger in a low-$Q^{2}$ region. The ratio $G_{E}^{p}/G_{M}^{p}$ can be extracted by measuring spin-dependent asymmetry from longitudinally 
polarized electrons scattering off from a polarized proton target~\cite{SANE:2018cub}. 
A recoil proton polarization measurement~\cite{JeffersonLabHallA:1999epl,JeffersonLabHallA:2001qqe,Gayou:2001qt,Zhan:2011ji,Puckett:2010ac,Puckett:2011xg}, 
in which a longitudinally polarized electron beam scatters off from an unpolarized proton target, and the recoil proton polarization is measured, provides 
another way to determine the proton electric-to-magnetic form-factor ratio. 
Several crucial experiments using the Rosenbluth technique have been conducted for measuring the proton form-factor ratio in a 
higher $Q^{2}$ range of $0.4 - 5.5~{\rm (GeV/c)^{2}}$ \cite{E94110:2004lsx,Qattan:2004ht,Christy:2021snt}, and for addressing its discrepancy
from the ratio measured with the polarization transfer technique.
By combining unpolarized cross-section measurements with the form-factor ratio determined from one of the two aforementioned methods, 
one can extract the two proton form factors without the limitation of the Rosenbluth separation method~\cite{Arrington:2007ux}.

In this paper, we consider two recent $e-p$ scattering experiments. The first experiment was conducted at the experimental setup of the A1 
collaboration \cite{A1:2010nsl,A1:2013fsc,Bernauer:2010zga,Blomqvist:1998xn} at Mainz Microtron (MAMI)
\cite{Herminghaus:1976mt,Kaiser:2008zza,Jankowiak:2006yc}, and the second one at the setup of the PRad collaboration at Jefferson Lab Hall 
B \cite{Gasparian:2011,Gasparian:2014rna,Meziane:2013yma,refId0,Xiong:2019umf,Xiong:2020kds} with the CEBAF accelerator \cite{Leemann:2001dg}.
The A1 detector setup consists of three high resolution magnetic spectrometers, labeled A, B and C, which can be rotated around the central 
axis to perform measurements at various scattering angles. The measured scattering angle is detected with an absolute accuracy 0.01$^{\circ}$ (0.175 mrad). 
The PRad setup utilizes a magnetic-spectrometer-free, calorimeter-based method along with a windowless hydrogen gas target, by means of which 
several limitations of the previous $e-p$ experiments have been overcome, allowing the setup to achieve the forward most scattering angle 
of $\sim 0.7^\circ$. Both experiments have extracted the proton electric form factor and the electric (charge) radius ($r_{E}^{p}$) from 
cross-section data analyzed in the ranges of $Q^{2} = 0.0038 - 0.9772~{\rm (GeV/c)^{2}}$ for Mainz A1 and $Q^{2} = 0.0002 - 0.0582~{\rm (GeV/c)^{2}}$ 
for PRad. 

We are particularly interested in the discrepancy of $G_{E}^{p}$ extracted by Mainz A1 and PRad. The $Q^{2}$-range measured in PRad has an
overlap with the Mainz data above $Q^{2} = 0.0038~{\rm (GeV/c)^{2}}$. In that range, $G_{E}^{p}$ obtained from fits to the Mainz and 
earlier data falls about 1.5\% faster than the PRad data, which is shown in Figure~2 of \cite{Bernauer:2020ont} for example. As pointed out in 
\cite{Bernauer:2020ont}, this discrepancy in the overlap $Q^{2}$ region cannot be explained by problems in the extrapolation to $Q^{2} = 0$, or 
by an inadequate selection of fit models (like inadequate fit functions or $Q^{2}$ ranges). In this regard, a few upcoming experiments, e.g.,
PRad-II \cite{PRad:2020oor} may solve this issue\footnote{The PRad-II experiment is designed to reduce the total uncertainty on $r_{E}^{p}$ by 
a factor of $\sim 4$ compared to that of PRad.}. On the other hand, one can still try to address the $G_{E}^{p}$ discrepancy by focusing on 
the last step of the form factor extraction -- the fitting. What we mean is explained in what follows.

PRad has developed an expandable framework for finding mathematical functions (fitters), which allows for extraction of $r_{E}^{p}$ in a 
robust way, by using pseudo-data generated over a broad set of various $G_{E}^{p}$ input functions \cite{Yan:2018bez, Xiong:2020kds}. This 
has been done using the expected $Q^{2}$ binning and uncertainties of the PRad experiment, prior to the extraction of $r_{E}^{p}$ from the 
PRad data on $G_{E}^{p}$. A similar method is also applied in the studies of \cite{A1:2013fsc} and \cite{Kraus:2014qua}.
The fact that different analytic choices can impact the extraction of $G_{E}^{p}$ from elastic $e-p$ cross-section 
data \cite{Barcus:2019skg} motivates a desire to use the lowest order (N=1, M=1), two-parameter fitting function, Rational\,(1,1) (applied by 
PRad in fitting its $G_{E}^{p}$ data and obtaining $r_{E}^{p}$), to fit the Mainz A1 data in some $Q^{2}$ range, and then compare this fit with 
the PRad data, as well as with the fit of the $\rm{10^{th}}$ order polynomial function (one of the functions applied by Mainz A1 in obtaining 
its $G_{E}^{p}$ and $r_{E}^{p}$).

Nonetheless, our basic motivation for using the lowest order rational function and a selected $Q^{2}$ range is anchored upon the results of
\cite{Alarcon:2017ivh,Alarcon:2017lhg,Alarcon:2018irp,Alarcon:2018zbz,Alarcon:2020kcz}. In particular, a novel theoretical framework 
called dispersively improved chiral effective field theory (DI$\chi$EFT) is introduced and developed in
\cite{Alarcon:2017ivh,Alarcon:2017lhg,Alarcon:2018irp} to calculate the nucleon form factors. The DI$\chi$EFT framework combines the chiral 
effective field theory with dispersion analysis, and produces theoretical parametrizations describing the nucleon electromagnetic form factors. 
The calculation with controlled uncertainties is based on the first principles of DI$\chi$EFT, and is consistent with empirical amplitude analysis 
\cite{Alarcon:2018irp}, which is 
a state-of-the-art theoretical calculation. 
In this regard, the Rational\,(1,1) function is simply taken by us as a reasonable approximation, which has been checked by 
comparing it to those extremely complex functions in \cite{Alarcon:2017ivh,Alarcon:2017lhg,Alarcon:2018irp}.
This framework, which is applicable up to $Q^{2} \approx 0.5~{\rm (GeV/c)^{2}}$, was used for extraction of the proton electric and magnetic 
radii from the world data and the Mainz A1 $e-p$ scattering data \cite{Alarcon:2018zbz,Alarcon:2020kcz}. 

In addition, one should note that the effect from the two-photon-exchange (TPE) correction has been investigated and found to be negligible
in the PRad kinematics. In particular, $e-p$ event generators used in the simulations for the PRad $r_{p}$ extraction \cite{Xiong:2020kds}, 
also included the contribution from the TPE processes (studied in \cite{Tomalak:2018ere,Tomalak:2015aoa,Tomalak:2014sva}), which was
estimated to be $< 0.2\%$ of the $e-p$ elastic scattering cross section in the PRad kinematic range. 
In a recent study, based on the nucleon form factors obtained within the dispersion theoretical framework in \cite{Lin:2021cnk},
the differential cross sections are calculated for $e-p$ and $e^{+}-p$ at the PRad/PRad-II kinematics for such electron and positron elastic
scattering experiments \cite{PRad:2020oor,Hague:2021xcc}. The cross section sensitivity to different sets of TPE corrections is investigated. This study show that an uncertainty emerging from those corrections becomes a secondary effect when it is compared with the nucleon form-factor 
uncertainties at the PRad/PRad-II beam energies. The TPE contribution has also been determined to be very small \cite{Blunden:2005jv} in an earlier study of $r_{p}$ determination. While the TPE effects are not significant in magnitude, they introduce a strong angular dependence in the cross section
\cite{Blunden:2003sp,Blunden:2005ev}, which contribute to the discrepancy in the proton form-factor ratio determined at high $Q^{2}$ between the
Rosenbluth and polarization transfer measurements.

Based upon using the Rational\,(1,1) fitter function, along with studying its comparison with DI$\chi$EFT form-factor parametrizations,
we perform a re-analysis of the Mainz A1 cross-section data with its 34 different combinations of the unconstrainted 31
normalization parameters up to $Q^{2} = 0.5~{\rm (GeV/c)^{2}}$, and show that this re-analysis possibly resolves the $G_{E}^{p}$ discrepancy 
puzzle between the Mainz A1 and PRad data. However, to be more specific, we use 31 sets of 29 normalization parameters within this $Q^{2}$ range.
The paper is organized as follows. Sec.~\ref{sec:formalism} introduces the notations and formalism pertinent to our discussion. 
In Sec.~\ref{sec:R11-theory}, we discuss the comparison of the DI$\chi$EFT parametrizations with the Rational\,(1,1), as well as the fit of 
Rational\,(1,1) to the Mainz A1 data. We show our main result in Sec.~\ref{sec:sum}, and discuss some would-be prospects.

\section{\label{sec:formalism} Notations and formalism}

If we focus on the Sachs electric and magnetic form factors of the proton, they are interpreted as Fourier transforms of the proton charge 
and magnetic moment distributions (in the non-relativistic limit), given by
\beq
G_{E/M}^{p}(Q^{2}) = \bigintssss \rho_{E/M}^{p}(\vec{x})\,e^{i \vec{q} \cdot \vec{x}}\,d^{3}x .
\label{eq:eqn_SachsFF}
\eeq
In the so-called Breit frame, where the proton bounces back after absorbing a virtual photon, by also having the $z$-axis pointing in the 
direction of the incident proton, the following formulas for the kinematic variables should be used:
\beq
E_{p} = E_{p^{\prime}},~~~~\nu = 0,~~~~\vec{p} = -\vec{p^{\prime}} = \frac{1}{2}\,\vec{q},~~~~Q^{2} = |\vec{q}|^{2} .
\label{eq:eqn_kinematics}
\eeq
Assuming the Taylor expansion at the limit of $Q^{2} \rightarrow 0$, along with spherically symmetric density distributions, the form factors 
are represented as
\bea
& & \!\!\!\! G_{E/M}^{p}(Q^{2}) = 
\nonumber \\
& & ~~ = \bigintssss \left( 1 + i \vec{q} \cdot \vec{x} - \frac{ \left(  \vec{q} \cdot \vec{x} \right)^{2} }{2} 
+ ...\right) \rho_{E/M}^{p}(\vec{x})\,d^{3}x =
\nonumber \\
& & ~~ = 1- \frac{1}{6}\,{\langle r_{E/M}^{p~~~2} \rangle}\,Q^{2} + C_{2}\,Q^{4} + ... ,
\label{eq:eqn_Taylor}
\eea
where the proton electric and magnetic 3-dimensional root-mean-square radii according to Sachs \cite{Sachs:1962zzc} are identified as
\bea
& & r_{E/M}^{p} \equiv r_{E/M,rms}^{p} \equiv \sqrt{\langle r_{E/M}^{p~~~2} \rangle} 
\nonumber \\
& & ~~~~~~~= \left( -\frac{6}{G_{E/M}^{p}(0)} \left. \frac{\mathrm{d} G_{E/M}^{p}(Q^2)} {\mathrm{d}Q^2} \right|_{Q^{2}=0} \right)^{1/2} ,
\label{eq:eqn_rp}
\eea
where $G_{E}^{p}(0) = 1$ and $G_{M}^{p}(0) = \mu^{p} = 2.7928\mu_{N}$ \cite{Schneider:2017lff}.

One should note that for a theory-based analysis, it is advantageous to use the Dirac, $F_{1}^{p}(Q^{2})$, and Pauli, $F_{2}^{p}(Q^{2})$, form 
factors of the proton \cite{Lin:2021umk}, which are related to the Sachs $Q^{2}$-dependent form factors by the following linear combinations:
\bea
& & G_{E}^{p}(Q^{2}) = F_{1}^{p}(Q^{2}) + \frac{Q^{2}}{4 m_{p}^{2}}\,F_{2}^{p}(Q^{2}) ,
\nonumber \\
& & G_{M}^{p}(Q^{2}) = F_{1}^{p}(Q^{2}) + F_{2}^{p}(Q^{2}) .
\label{eq:eqn_FFs}
\eea

In one-photon exchange approximation, the differential Born cross section for the elastic $e-p$ scattering is given in terms of the Sachs 
$Q^{2}$-dependent form factors $G_{E}^{p}(Q^{2}) \equiv G_{E}^{p}$ and $G_{M}^{p}(Q^{2}) \equiv G_{M}^{p}$~\cite{Hand:1963zz}.
%
%
\bea
& & \!\!\!\!\!\!\!\!
\Lb \frac{d\sigma}{d\Omega} \Rb_{0} = \Lb \frac{d\sigma}{d\Omega} \Rb_{\rm Mott} \times
\nonumber \\
& & \times \left( \frac{(G_{E}^{p})^{2} + \tau\,(G_{M}^{p})^{2}}{1 + \tau} + 2\tau\,(G_{M}^{p})^{2}\tan^{2}{\!\Lb \frac{\theta}{2} \Rb} \right) =
\nonumber \\
& & = \Lb \frac{d\sigma}{d\Omega} \Rb_{\rm Mott} \frac{\varepsilon\,(G_{E}^{p})^{2} + \tau\,(G_{M}^{p})^{2}}{\varepsilon\,(1 + \tau)} ,
\label{eq:eqn_sigma}
\eea
where $\Lb d\sigma/d\Omega \Rb_{\rm Mott}$ is the recoil-corrected relativistic Mott cross section on a point-like particle:
\beq
\Lb \frac{d\sigma}{d\Omega} \Rb_{\rm Mott} = 4\alpha^{2} \cos^{2}{\!\Lb \frac{\theta}{2} \Rb}\,\frac{1}{Q^{4}} \frac{(E^{\prime})^{3}}{E} .
\label{eq:eqn_Mott}
\eeq
The dimensionless kinematic variables $\varepsilon$ and $\tau$ are given by 
\beq
\varepsilon = \Lb 1 + 2(1 + \tau) \tan^{2}{\!\Lb \frac{\theta}{2} \Rb} \Rb^{-1} ,~~~~~~~\tau = \frac{Q^{2}}{4m_{p}^{2}} ,
\label{eq:eqn_dless}
\eeq
and the four-momentum transfer squared is 
\beq
Q^{2} = -q^{2} = 4 E E^{\prime} \sin^{2}{\!\Lb \frac{\theta}{2} \Rb} ,
\label{eq:eqn_4mom}
\eeq
describing an electron of energy $E$, scattering off a proton at rest through an angle $\theta$ with respect to the beam direction of the 
scattered electron of energy $E^{\prime}$ that is expressed as
\beq
E^{\prime} = E \Lb 1 + \frac{2E}{m_{p}} \sin^{2}{\!\Lb \frac{\theta}{2} \Rb} \Rb^{-1} .
\label{eq:eqn_Escat}
\eeq

The deduction of the Born cross section shown in Eq.~(\ref{eq:eqn_sigma}) requires a correction factor, $f_{\rm corr}$, from calculations of 
radiative and other effects. Therefore, the relation of the experimental and Born cross sections can be written as
\beq
\Lb \frac{d\sigma}{d\Omega} \Rb_{\rm exp} =  f_{\rm corr} \Lb \frac{d\sigma}{d\Omega} \Rb_{0}.
\label{eq:eqn_ExpBorn}
\eeq

The published elastic $e-p$ scattering cross-section data from the Mainz A1 experiment includes 34 sets of data and 31 independent 
normalization factors as discussed in \cite{A1:2010nsl,A1:2013fsc,Bernauer:2010zga}. The data is presented as the cross-section ratio to 
the dipole cross section, which is simply Eq.~(\ref{eq:eqn_sigma}) but with the form factors there taken to be the standard dipole ones:
\bea
& & G_{E,{\rm dipole}}^{p}(Q^{2})  = \Lb 1 + \frac{Q^{2}}{0.71~{\rm (GeV/c)^{2}}} \Rb^{-2} ,
\nonumber \\
& & G_{M,{\rm dipole}}^{p}(Q^{2}) = \mu_{p}\Lb 1 + \frac{Q^{2}}{0.71~{\rm (GeV/c)^{2}}} \Rb^{-2} ,
\label{eq:eqn_dipole}
\eea
and that ratio (after all corrections) is expressed as
\bea
& & \!\!\!\!\!\!\!\!\!\!\! 
\frac{\sigma_{\rm exp}}{\sigma_{\rm dipole}} \equiv \Lb \frac{\varepsilon\,(G_{E}^{p})^{2} + \tau\,(G_{M}^{p})^{2}}{\varepsilon\,(G_{E,{\rm dipole}}^{p})^{2} 
+ \tau\,\mu_{p}^{2}\,(G_{E,{\rm dipole}}^{p})^{2}} \Rb .
\label{eq:eqn_ratio}
\eea

In order to obtain the final cross-section ratio values, 34 groups of various cross-section ratios are multiplied by the 31 independent 
normalization factors according to
\beq
\frac{\sigma_{\rm exp}}{\sigma_{\rm dipole}} \rightarrow n_{1} \times n_{2} \times \frac{\sigma_{\rm exp}}{\sigma_{\rm dipole}} ,
\label{eq:eqn_xs_norm}
\eeq
where $n_1$ and $n_2$ are any pair of the 31 normalization factors, and any such combination for each group of data points is 
given in the supplemental material of \cite{A1:2013fsc}. In the analysis herein, we use only the data sets less than or equal to
$Q^{2} = 0.5~{\rm (GeV/c)^{2}}$, and thus only 29 independent normalization parameters.

\section{\label{sec:R11-theory} An analysis based on using the lowest order rational function motivated by the
\texorpdfstring{DI$\chi$EFT}{DIchiEFT} calculations}
\subsection{\label{subsec:R11_DIXEFT} Comparison of the Rational\texorpdfstring{\,(1,1)}{(1,1)} to the \texorpdfstring{DI$\chi$EFT}{DIchiEFT} parametrizations}

The DI$\chi$EFT theoretically-derived parametrizations of the proton electromagnetic form factors
\cite{Alarcon:2017ivh,Alarcon:2017lhg,Alarcon:2018irp} can be expressed in terms of the nucleon electric and magnetic radii
(see the definition in Eq.~(\ref{eq:eqn_rp})):
\bea
& & \!\!\!\!\!\!\!\! G_{E}^{p}(Q^2) = 
\nonumber \\
& & ~~= A_{E}^{p}(Q^2)+\langle r_{E}^{p~2} \rangle B_{E}(Q^2)+\langle r_{E}^{n~2} \rangle\bar{B}_{E}(Q^2) ,
\nonumber \\
& & \!\!\!\!\!\!\!\! G_{M}^{p}(Q^2) = 
\nonumber \\
& & ~~= A_{M}^{p}(Q^2)+\langle r_{M}^{p~2} \rangle B_{M}(Q^2)+\langle r_{M}^{n~2} \rangle\bar{B}_{M}(Q^2) ,
\label{eq:eqn_DIXEFT_FFs}
\eea
where the functions $A_{E/M}^{p}(Q^2)$ are the radius-independent parts, while the functions $B_{E/M}(Q^2)$, $\bar{B}_{E/M}(Q^2)$ 
are the parts proportional to the nucleon electric/magnetic radii. The details of the parametrizations shown in 
Eq.~(\ref{eq:eqn_DIXEFT_FFs}) and the values of these functions at different $Q^2$ points from $0$ to $\rm{1~(GeV/c)^2}$ 
are given in the supplemental materials of \cite{Alarcon:2020kcz}. 
 
So, DIXEFT employs fixed radius values as its input. In the analysis of the Mainz data using the DIXEFT method
\cite{Alarcon:2020kcz}, Alarcon {\it et.al} have used different radii as input in the DIXEFT calculation and compared it with
the Mainz data to obtain the most overlapping radii. This method is different from the methods used in the Mainz and PRad
analyses, which use analytic functions to fit the data and carry out the extrapolation to derive $r^p_E$ directly.

As discussed in \cite{Barcus:2019skg}, the choice of an analytic function has a dramatic impact on the form-factor extraction. 
The results of this paper demonstrate that when using regression algorithms, it is crucial to have a solid mathematical basis and/or a good 
physical model. Consequently, it is essential for a selected analytic function to have a strong physical meaning, such that it can be well 
compared to theoretical calculations. Generally, one can think of statistical predictive models, which attempt to generalize beyond some data 
sets that are being fitted. Such models require incorporation of physics considerations and/or mathematical requirements to keep the fits
well-behaved, rather than making them of intricate nature.

In this section, we make use of the Rational\,(1,1), which is determined in PRad studies to be the best robust fitter for extraction 
of the proton electric radius~\cite{Xiong:2019umf,Xiong:2020kds,Yan:2018bez}. This functional form is given by
\beq
f_{\rm Rational\,(1,1)}(Q^{2}) \equiv {\rm Rational\,(1,1)} = \frac{1 + p_{1}^{(a)}Q^{2}}{1 + p_{1}^{(b)}Q^{2}} ,
\label{eq:eqn_R11}
\eeq
where the coefficients $p_{1}^{(a)}$ and $p_{1}^{(b)}$ are two free fitting parameters\footnote{We borrow the same nomenclature as the one used 
in \cite{Yan:2018bez,Zhou:2020cdt}.}. The rational function in Eq.~(\ref{eq:eqn_R11}) is the lowest order of the multi-parameter rational-function 
of $Q^{2}$ represented by 
\bea
& & \!\!\!\!\! f_{\rm Rational\,(N,M)}(Q^{2}) \equiv {\rm Rational\,(N,M)} = 
\nonumber \\
& & ~~~= \left( 1 + \sum\limits_{i=1}^{\rm N} p_{i}^{(a)}Q^{2i} \right)~\Bigg/~\left( 1 + \sum\limits_{j=1}^{\rm M} p_{j}^{(b)}Q^{2j} \right) ,
\label{eq:eqn_RNM}
\eea


The form factors $G_{E}^{p}$ and $G_{M}^{p}$ given in terms of the Rational\,(1,1) will be represented as
\beq
\!\! G_{E}^{p}(Q^{2}) = \frac{1 + p_{1,E}^{(a)} Q^{2}}{1 + p_{1,E}^{(b)} Q^{2}} ,~
G_{M}^{p}(Q^{2}) = \mu_{p} \frac{1 + p_{1,M}^{(a)} Q^{2}}{1 + p_{1,M}^{(b)} Q^{2}} .
\label{eq:eqn_R11_GEGM}
\eeq
The fitting parameters have a direct relation to the proton electric/magnetic radii, which is expressed by 
\beq
r^{p}_{E/M} = \sqrt{6 \left(p_{1,E/M}^{(b)} - p_{1,E/M}^{(a)} \right)} . 
\eeq

In order to compare the Rational\,(1,1) with the DI$\chi$EFT parametrizations, we use Eq.~(\ref{eq:eqn_DIXEFT_FFs}) to generate a set of 
$G_{E/M}^{p}$ pseudo-data at various equidistant $Q^{2}$ points in the range of $0$ to $\rm{1~(GeV/c)^2}$. Then we fit the Rational\,(1,1) to 
the pseudo-data in different ranges of $Q^2$ from 0 to $Q^2_{\rm max}$, respectively\footnote{$Q_{\rm max}^{2}$ is an upper value in a specific 
$Q^{2}$ binning set, which we take from the full range of the Mainz A1 cross-section data set.}. In the pseudo-data generation, the neutron 
electromagnetic radii are fixed to be $\langle r_{E}^{n~2} \rangle = -0.116~{\rm fm^{2}}$ and $r^{n}_{M} = 0.864~{\rm fm}$, 
whereas the proton radii $r^p_{E/M}$ are varied in the region from $0.80~{\rm fm}$ to $0.88~{\rm fm}$. After obtaining the fitting parameters 
of Rational\,(1,1) from the fit, the residual can be calculated according to the formula shown below, which characterizes the difference 
between the fitted Rational\,(1,1) and the DI$\chi$EFT parametrizations defined as
\bea
& & \!\!\!\!\!\!\!\!\! G_{E/M}^{p}~\rm{Residual} \equiv 
\nonumber \\
& & \!\!\!\!\!\!\!\!\! \frac{\sum\limits^{N_{\rm points}} \Lb G^{p,\rm ~Rational\,(1,1)}_{E/M}(Q^2) - 
G^{p, \rm ~DI \chi EFT}_{E/M}(Q^2)\Rb^{2}}{N_{\rm points}} ,
\label{eq:eqn_residual}
\eea
where $N_{\rm points}$ is the number of points within the $Q^2$ range.
\begin{figure}[hbt!]
\centering
\includegraphics[width=8.0cm]{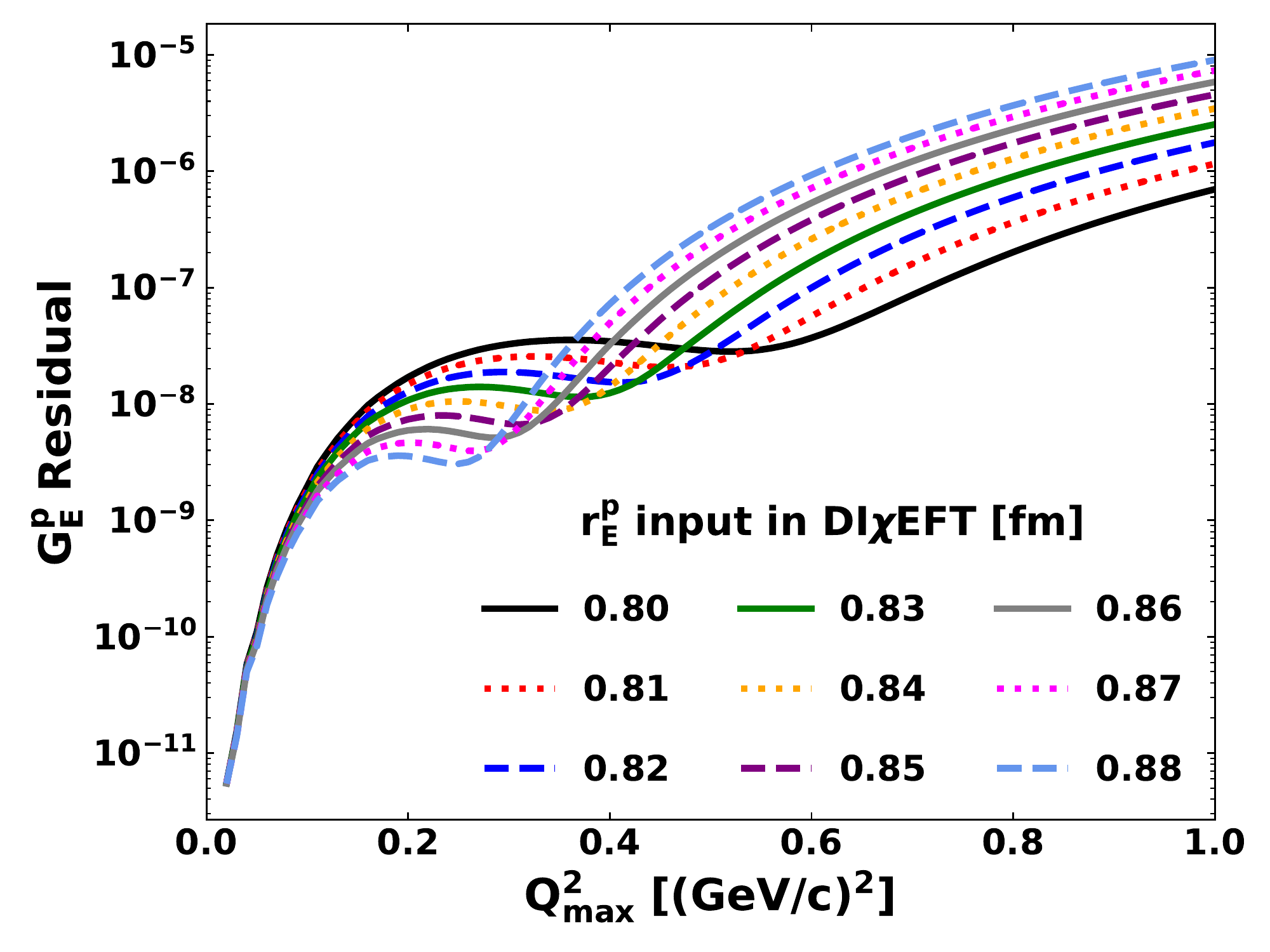}
\includegraphics[width=8.0cm]{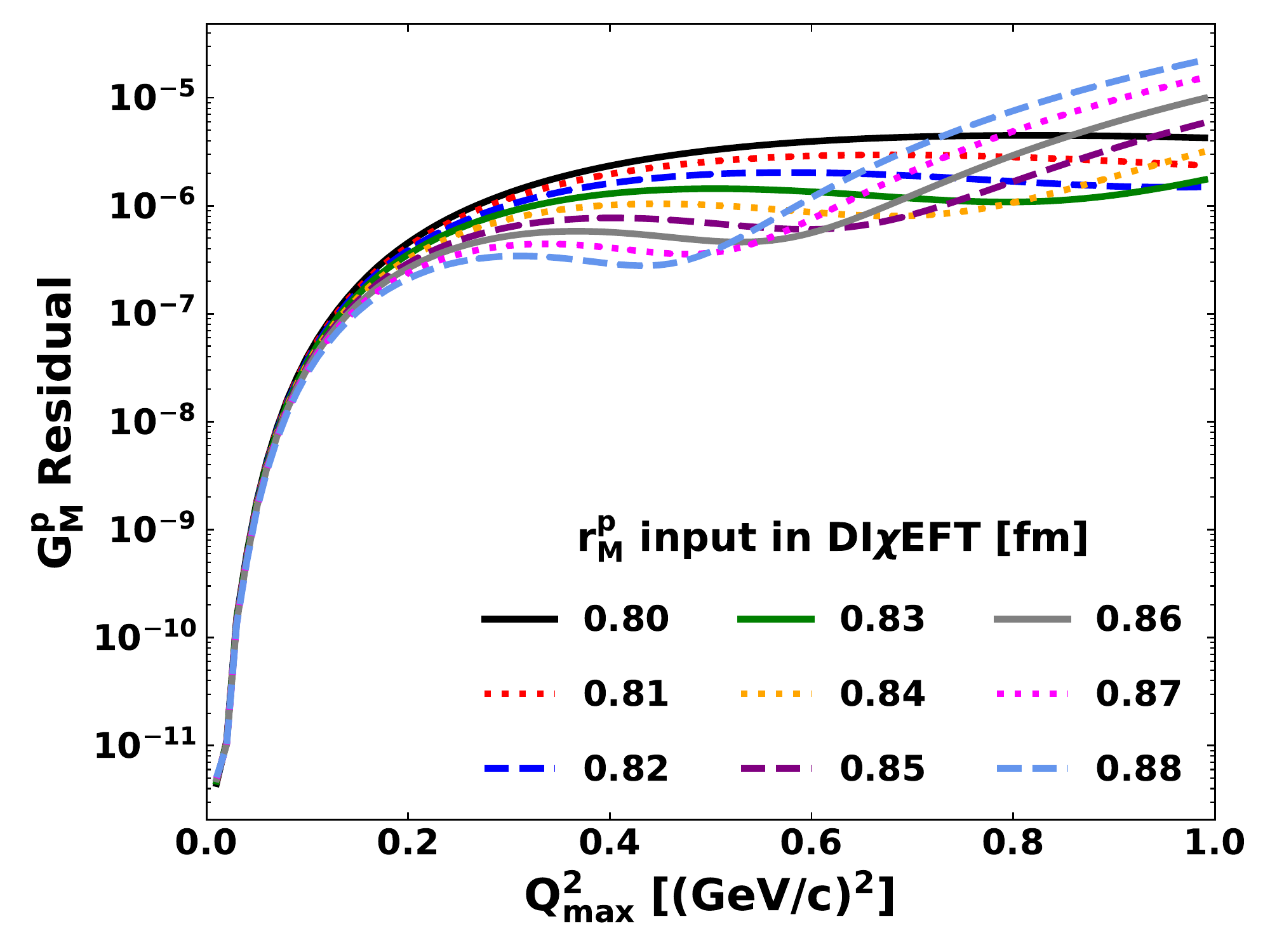}
\caption{(Color online) $G_{E}^{p}$ (top) and $G_{M}^{p}$ (bottom) residuals between the fitted Rational\,(1,1) 
and the DI$\chi$EFT parametrizations as a function of $Q^2_{\rm max}$ (end point of the fit). The neutron electromagnetic radii are 
fixed to be $\langle r_{E}^{n~2} \rangle = -0.116~{\rm fm^{2}}$ and $r^{n}_{M} = 0.864~{\rm fm}$ \cite{ParticleDataGroup:2018ovx}, 
while the proton radii $r^p_{E/M}$ 
are varied from $0.80~{\rm fm}$ to $0.88~{\rm fm}$.}
\label{fig:Residual}
\end{figure}

The results of $G_{E}^{p}$ and $G_{M}^{p}$ residuals are shown in Fig.~\ref{fig:Residual}. In both cases, the residuals are small. 
Fig.~\ref{fig:Residual} shows that the Rational\,(1,1) has an excellent agreement with the DI$\chi$EFT parametrizations at low-$Q^2$ 
region. Besides, this result supports such a rational fitting function to be a good candidate for the robust and precise extraction 
of the proton electric radius, which has in turn been demonstrated in \cite{Xiong:2019umf,Xiong:2020kds}.

\subsection{\label{subsec:Fitting} Fitting of the Rational\texorpdfstring{\,(1,1)}{(1,1)} to the Mainz A1 cross-section data}

In the Mainz A1 data set, there are 1422 cross-section data points in the entire range of $Q^{2}$ from 0.0038 to 0.9772 ${\rm (GeV/c)^{2}}$. 
Note that there are $538$ data points in the overlap $Q^{2}$ region with the PRad data set, and this overlap region is
$Q^{2} = 0.0038 - 0.0582~{\rm (GeV/c)^{2}}$. Also, for the new regressions done herein with the Rational\,(1,1) function,
over a range of $Q^2$ from 0 to 0.5~${\rm (GeV/c)^{2}}$, we are using 90\% of the Mainz data, 1285 out of 1422 cross-section points. 
This is simply the range, where we have validated the function against the theory, and the presented results do not significantly change 
if the entire set is used.

Based upon the discussion hitherto, we assume that $G_{E}^{p}$ and $G_{M}^{p}$ have the forms of Rational\,(1,1) 
(see Eq.~(\ref{eq:eqn_R11_GEGM})).
We plug Eq.~(\ref{eq:eqn_R11_GEGM}) into Eq.~(\ref{eq:eqn_ratio}), then fit it to the Mainz A1 cross-section data up to 
$Q_{\rm max}^{2} = 0.5~{\rm (GeV/c)^{2}}$. In this fitting range, there are 31 sets of data with 29 independent normalization factors. 
We then use 29 independent normalization factors in the fitting, which are multiplied by the fitting function according to Eq.~(\ref{eq:eqn_xs_norm}).
\begin{table}[ht]
\begin{tabular}{cccccc}
    \hline
Data set & $n_1$ & $n_2$ & Unbounded-poly & Bounded-poly & Rational (1,1) \\
    \hline
1        & 3     & -     & 0.9996              & 1.0032            & 0.9936        \\
2        & 1     & 3     & 0.9997              & 1.0020            & 0.9961        \\
3        & 1     & 4     & 0.9995              & 1.0037            & 0.9921        \\
4        & 1     & 5     & 0.9996              & 1.0030            & 0.9949        \\
5        & 2     & 4     & 0.9996              & 1.0013            & 0.9987        \\
6        & 2     & 5     & 0.9997              & 1.0005            & 1.0015        \\
7        & 9     & -     & 0.9996              & 1.0043            & 0.9914        \\
8        & 7     & 9     & 0.9996              & 1.0042            & 0.9915        \\
9        & 6     & 9     & 0.9996              & 1.0044            & 0.9912        \\
10       & 8     & 9     & 0.9999              & 1.0053            & 0.9932        \\
11       & 13    & -     & 0.9993              & 1.0041            & 0.9905        \\
12       & 14    & -     & 0.9992              & 1.0038            & 0.9900        \\
13       & 11    & 13    & 0.9993              & 1.0042            & 0.9906        \\
14       & 10    & 13    & 0.9996              & 1.0048            & 0.9902        \\
15       & 10    & 14    & 0.9996              & 1.0045            & 0.9897        \\
16       & 10    & 15    & 0.9994              & 1.0036            & 0.9917        \\
17       & 12    & 15    & 0.9990              & 1.0034            & 0.9877        \\
18       & 18    & -     & 0.9994              & 1.0045            & 0.9894        \\
19       & 19    & -     & 0.9993              & 1.0037            & 0.9897        \\
20       & 16    & 18    & 0.9995              & 1.0049            & 0.9903        \\
21       & 16    & 19    & 0.9995              & 1.0041            & 0.9906        \\
22       & 16    & 20    & 0.9993              & 1.0039            & 0.9916        \\
23       & 17    & 20    & 0.9993              & 1.0045            & 0.9891        \\
24       & 25    & -     & 0.9992              & 1.0042            & 0.9881        \\
25       & 21    & 25    & 0.9993              & 1.0037            & 0.9893        \\
26       & 21    & 26    & 0.9995              & 1.0048            & 0.9893        \\
27       & 23    & 26    & 0.9995              & 1.0051            & 0.9879        \\
28       & 22    & 26    & 0.9988              & 1.0042            & 0.9979        \\
29       & 29    & 30    & 0.9994              & 1.0038            & 0.9919        \\
30       & 27    & 29    & 0.9994              & 1.0043            & 0.9887        \\
31       & 27    & 31    & 1.0001              & 1.0039            & 0.9912        \\  
\hline
\end{tabular}
      \caption{The 29 independent normalization factors from the Rational\,(1,1) fit are multiplied 
      into combined factors for the 31 data sets, as well as into analogous combined factors from 
      the unbounded/bounded (with alternating signs) $\rm{10^{th}}$ order polynomial function fits discussed in
      \cite{Barcus:2019skg}.}
\label{tab:table_norm}
\end{table}

\begin{figure}[ht!]
\centering
\includegraphics[width=8.5cm]{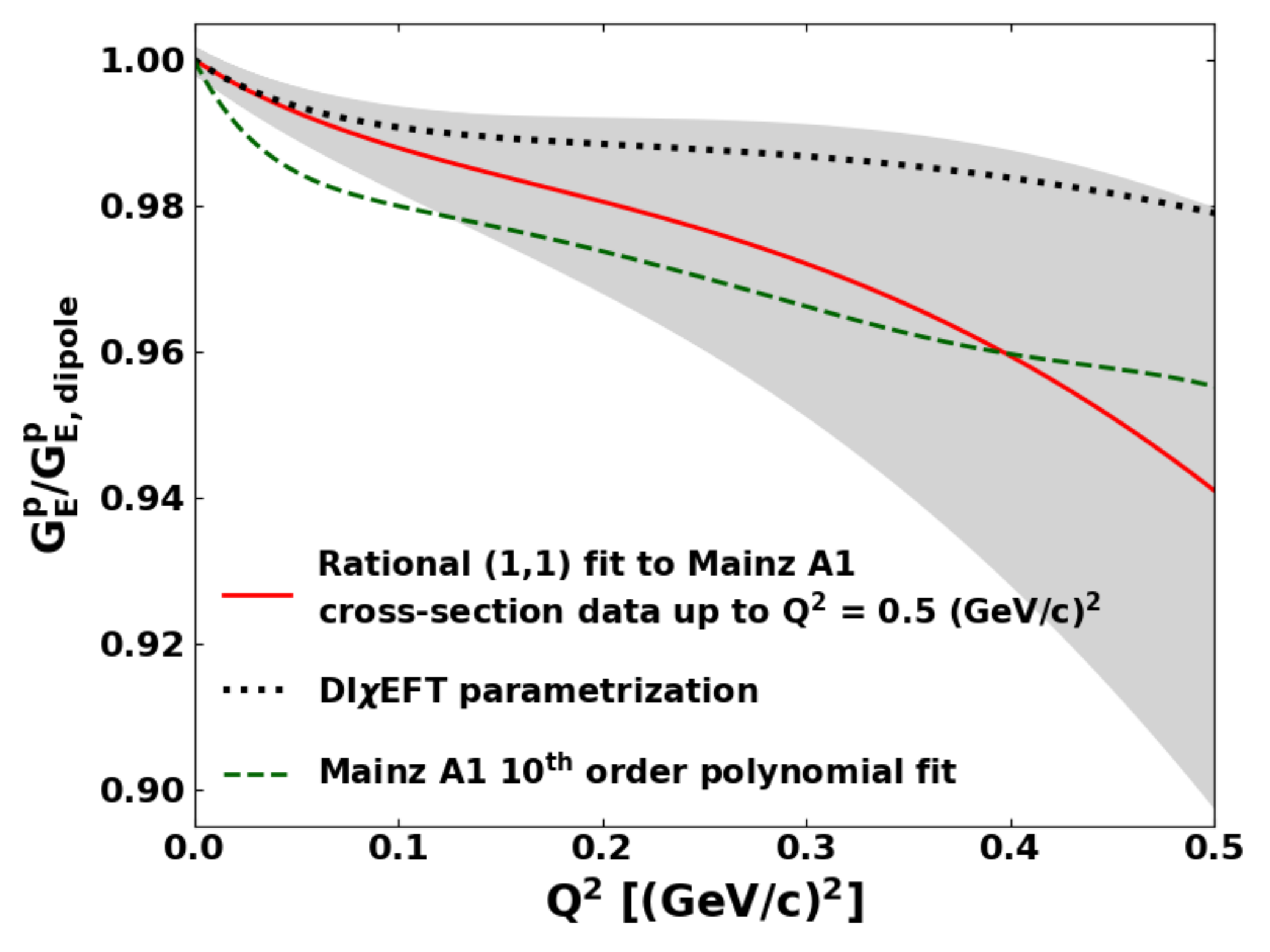}
\includegraphics[width=8.5cm]{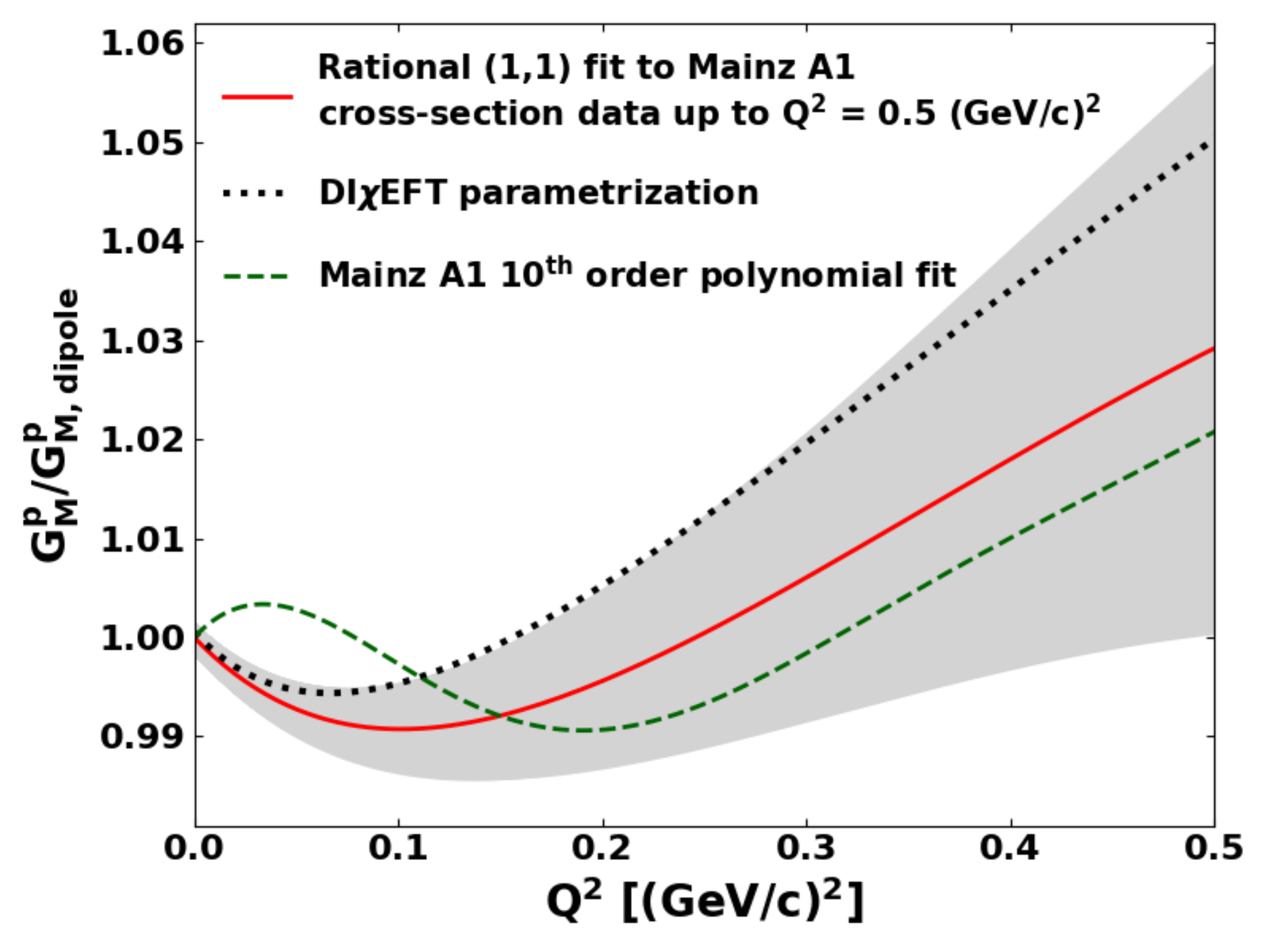}
\caption{(Color online) The fit results for $G_{E}^p$ (top) and $G_{M}^p$ (bottom) divided by the standard dipole 
form factors shown in Eq.~(\ref{eq:eqn_dipole}). The red solid curves show the Rational\,(1,1) best fit to the Mainz A1 cross-section data 
up to $Q_{\rm max}^{2} = 0.5~{\rm (GeV/c)^{2}}$; the grey bands show the 1$\sigma$ uncertainty of the Rational\,(1,1) fit; the green dashed 
curves show the $\rm{10^{th}}$ order polynomial function fitted to the Mainz A1 data taken from the supplemental material of 
\cite{A1:2013fsc}; the black dotted curves show the DI$\chi$EFT parametrizations.}
\label{fig:fit_GE_GM}
\end{figure}
The results of the extracted $G_{E}^{p}$ and $G_{M}^{p}$ divided by the standard dipole form factors from Eq.~(\ref{eq:eqn_dipole}) are shown 
in Fig.~\ref{fig:fit_GE_GM}, where the red solid curves are the Rational\,(1,1) best fits, and the grey bands show their 1$\sigma$ uncertainty 
regions. For comparison, we also plot the DI$\chi$EFT parametrizations, as well as the $\rm{10^{th}}$ order polynomial function 
fit taken from the supplemental material of \cite{A1:2013fsc}. It is shown that the wiggling behavior in $G_{M}^{p}$ from the 
$\rm{10^{th}}$ order polynomial function fit can be resolved by using the Rational\,(1,1) fit.

In the DI$\chi$EFT parametrizations, the neutron electromagnetic radii are fixed as $\langle r_{E}^{n~2} \rangle = -0.116~{\rm fm^{2}}$ and 
$r^{n}_{M} = 0.864~{\rm fm}$, which are average values taken from PDG \cite{ParticleDataGroup:2018ovx}. 
For the proton electromagnetic radii we take the central values obtained from the Rational\,(1,1) best fit: 
$r^{p}_{E} = 0.837~{\rm fm}$ and $r^{p}_{M} = 0.842~{\rm fm}$ with $\rm{\chi^2/dof} = 1.41$.
The $r^{p}_{E}$ result is consistent with several other re-analyses of this 
data~\cite{Mart:2013gfa,Lorenz:2014vha,Griffioen:2015hta,Higinbotham:2015rja,Alarcon:2018zbz,Horbatsch:2016ilr,Zhou:2018bon,Cui:2021vgm,Lin:2021umk} 
though inconsistent with some other re-analyses~\cite{Lee:2015jqa,Gramolin:2021gln}.
Differences between these fits, which are all fitting the same experimental data, can be traced back to  sub-percent level variations in the 29 
normalization parameters.

The corresponding 29 normalization factors from the Rational\,(1,1) fit are multiplied into combined factors for the 31 data sets as shown in 
Table.\ref{tab:table_norm}. For comparison, the combined factors from the unbounded and bounded $\rm{11^{th}}$ order polynomial function fits 
discussed in \cite{Barcus:2019skg} are also shown.  As shown, these normalization are in general consistent at the level of $1\%$ or less.

\begin{figure*}[hbt!]
\centering
\includegraphics[width=8.5cm]{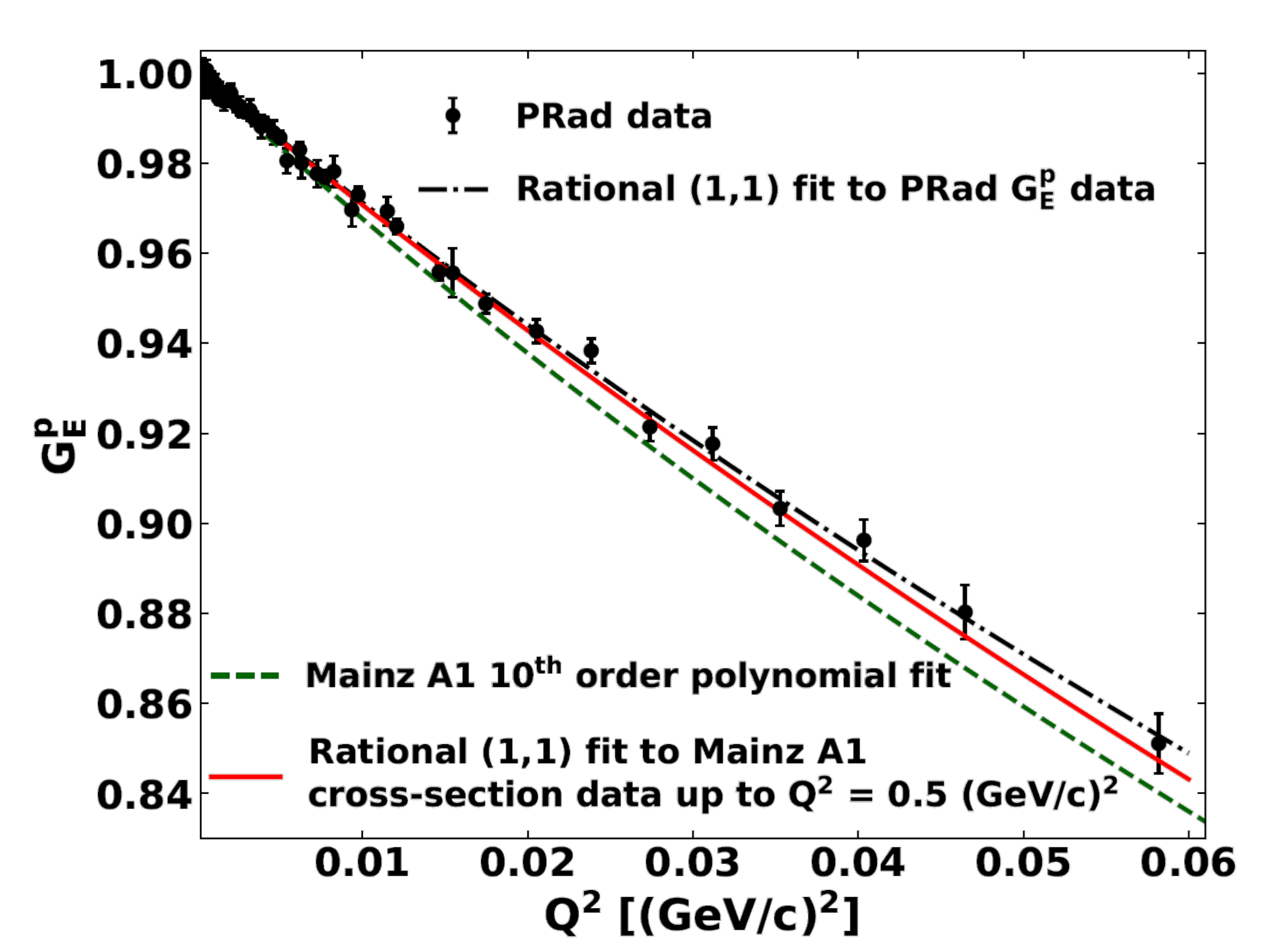}
\includegraphics[width=8.5cm]{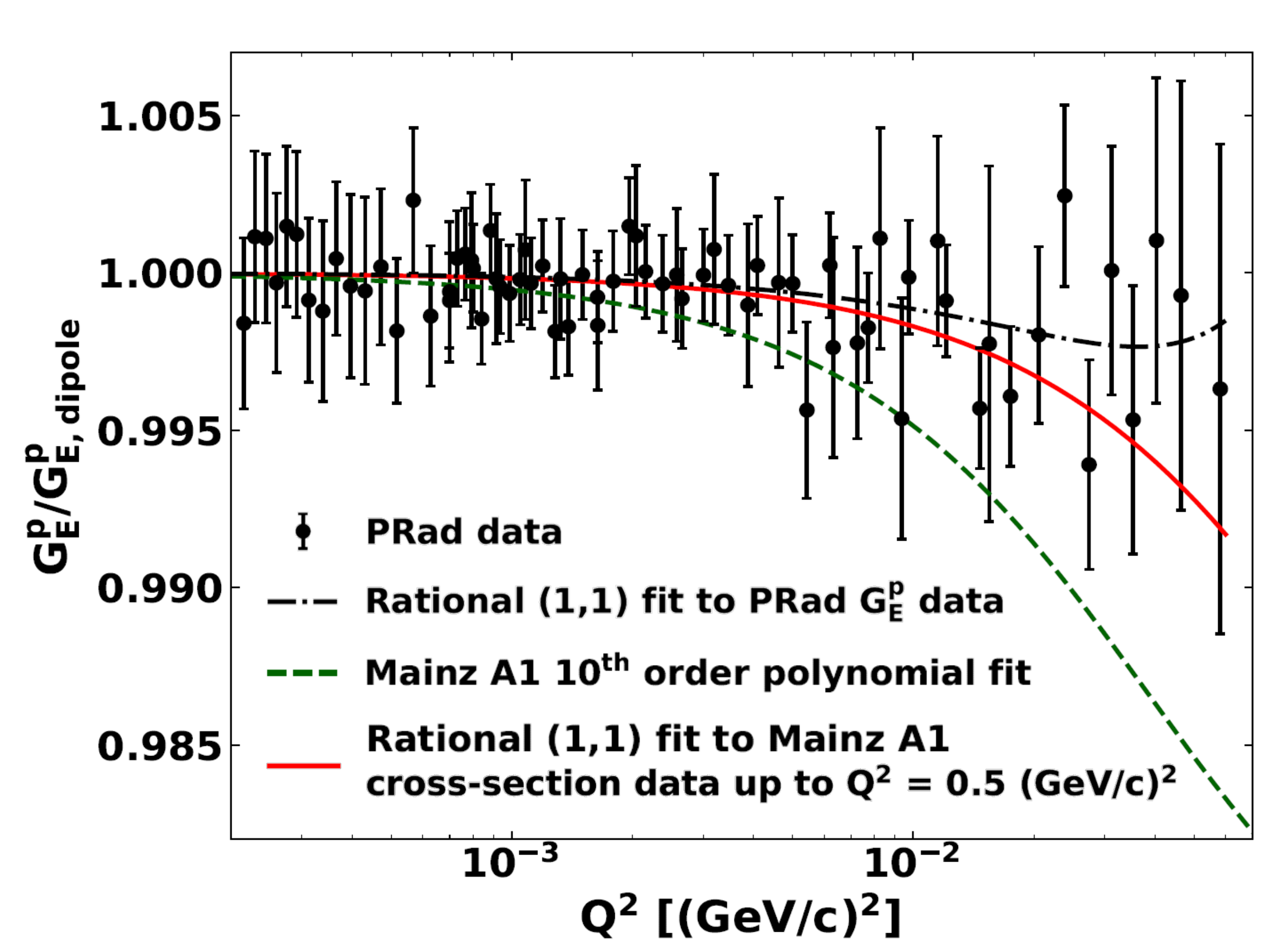}
\caption{(Color online) 
Comparison between the Rational\,(1,1) fit to the Mainz A1 cross-section data up to $Q^{2} = 0.5~{\rm (GeV/c)^{2}}$ and the $\rm{10^{th}}$ 
order polynomial function fit to the A1 data taken from the supplemental material of \cite{A1:2013fsc}. The PRad $G_{E}^{p}$ data 
with the statistical and systematic uncertainties summed in quadrature and the Rational\,(1,1) fit to the PRad data are shown as well.
The difference between the left and right plots is that in the right one we divide $G_{E}^{p}$ by the standard electric dipole 
parametrization from Eq.~(\ref{eq:eqn_dipole}).
}
\label{fig:final_compare}
\end{figure*}
\begin{figure*}[hbt!]
\centering
\includegraphics[width=8.75cm]{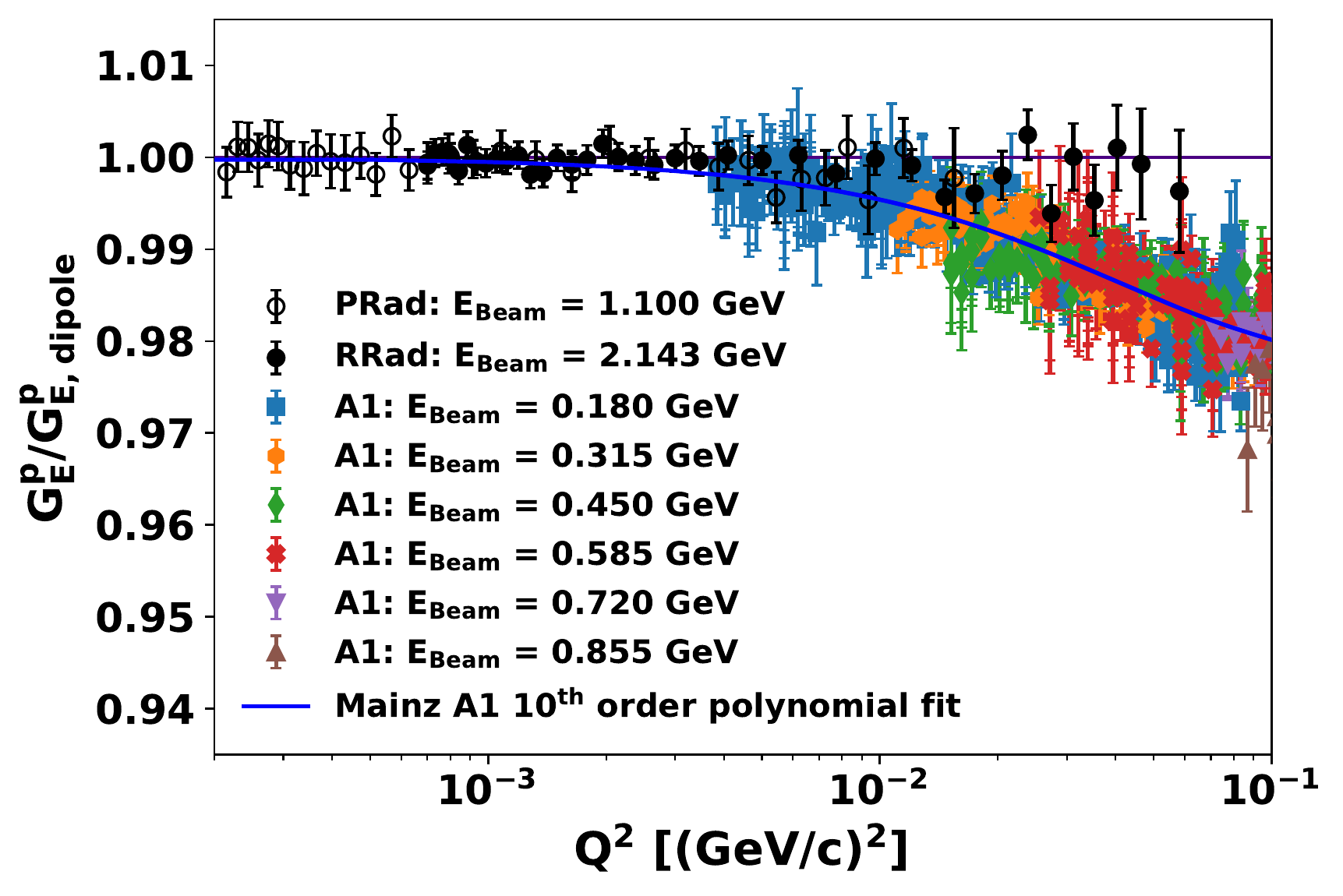}
\includegraphics[width=8.75cm]{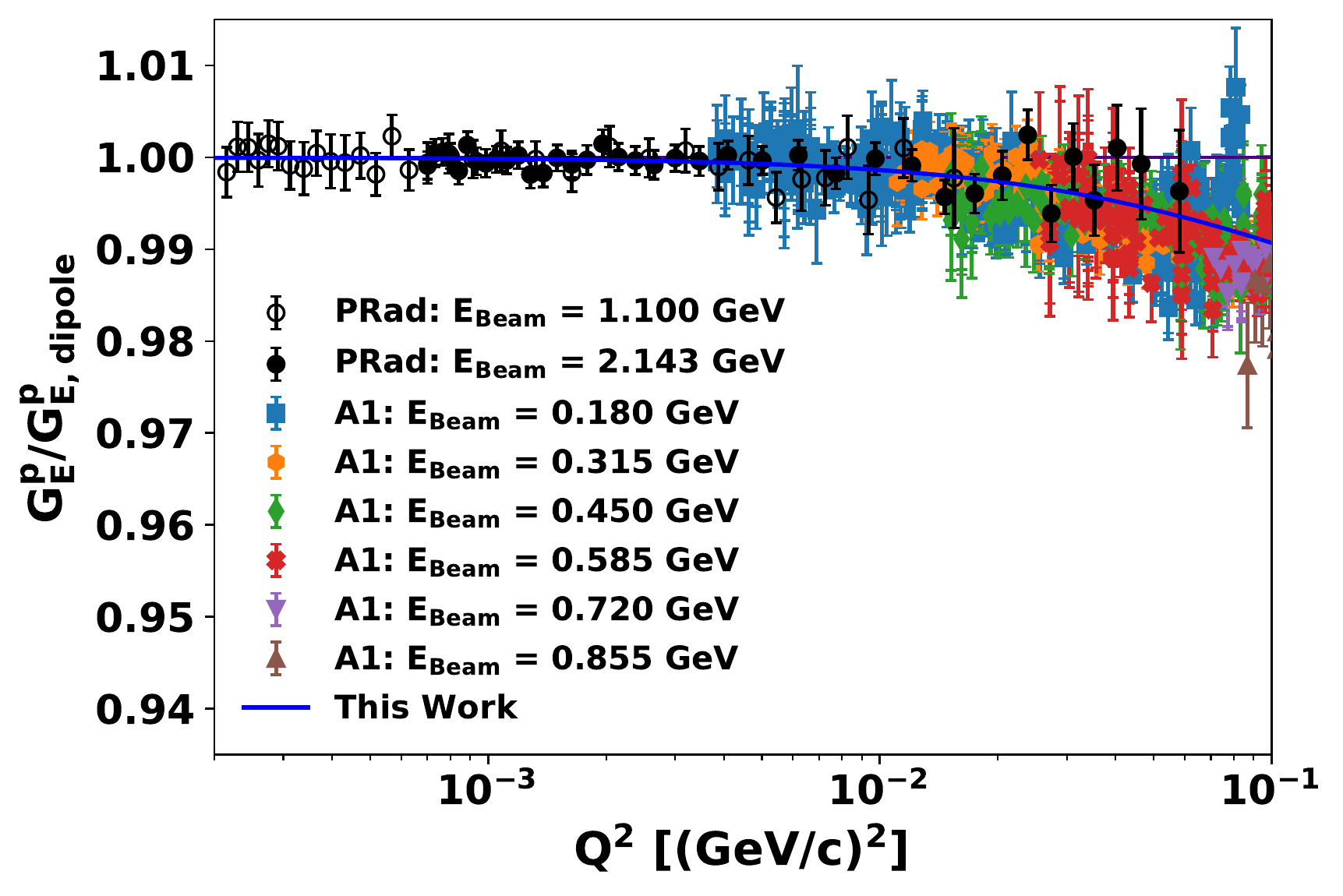}
\caption{(Color online) 
The left plot shows the $G_{E}^{p}$ form factor data extracted from the PRad and A1 experiments~\cite{A1:2010nsl,A1:2013fsc,Xiong:2019umf}. 
For the shown A1, the $G_{E}^{p}$ points are extracted from the cross sections using the floating normalizations driven 
by two $\rm{10^{th}}$ order polynomial functions used in a regression~\cite{A1:2013fsc}. The right plot shows the Rational\,(1,1) fit to the same 
data up to $Q^{2} = 0.5~{\rm (GeV/c)^{2}}$. Here the A1 cross-section normalization factors are determined with the Rational\,(1,1) fit.   
The uncertainties of the former are the statistical and systematic ones summed in quadrature. The horizontal lines are shown for the purpose of 
reference. This pair of plots clearly shows the influence the choice of a fit function can have on the A1 experiment's floating normalization parameters.
}
\label{fig:final_compare_norm}
\end{figure*}

\section{\label{sec:sum} Discussion and conclusion}

In this section we make use of validation data to see how well these fits generalize to data that have not been included in the regressions.
For this purpose we make use of the high precision, low $Q^2$ PRad data~\cite{Xiong:2019umf,Xiong:2020kds}.  
In Fig.~\ref{fig:final_compare}, which covers the $Q^2$ range of the PRad experiment, we again see the red solid curve describing the Rational\,(1,1) 
fit to the Mainz A1 cross-section data up to $Q^{2} = 0.5~{\rm (GeV/c)^{2}}$, and the green dashed curve describing the $\rm{10^{th}}$ order 
polynomial function fitted to the same Mainz A1 data set \cite{A1:2013fsc}. Also shown is the PRad data along with a  black dash-dotted curve 
which is a  Rational\,(1,1) fit to just PRad $G_{E}^{p}$ data \cite{Xiong:2019umf,Xiong:2020kds}. 

The proton electric form-factor discrepancy puzzle observed between the A1 and PRad data is visualized as the difference between 
the dashed and dash-dotted curves in the left and right plots of Fig.~\ref{fig:final_compare}. Thereby, one may consider 
the solid curve there to be a possible solution to this problem. Furthermore, we show in Fig.~\ref{fig:final_compare_norm} the comparison 
between the extracted $G_{E}^{p}$ data, using the Rational\,(1,1) fit to the A1 cross-section data up to $Q^{2} = 0.5~{\rm (GeV/c)^{2}}$ 
and the $\rm{10^{th}}$ order polynomial function fit to the same data taken from the supplemental material of~\cite{A1:2013fsc}. 
That figure clearly shows how the choice of a fit function is capable of shifting the floating normalization factor, and thus shifting the 
extracted form factors significantly. Here, by simply using the analytically well-behaved Rational\,(1,1) fit function, the extracted A1 
$G_{E}^{p}$ form factors become consistent with the PRad $G_{E}^{p}$ data within uncertainties.

However, it has been shown that changes in $e-p$ scattering data binning choices or in the choice of fitting function can result in substantially 
incompatible results~\cite{Barcus:2019skg}. As a consequence, we do not exclude a possibility that one can improve the results demonstrated in 
Fig.~\ref{fig:final_compare} and Fig.~\ref{fig:final_compare_norm} by employing an adequate selection of other fit models. It is perhaps 
plausible to construct such models with constraints, such as the Rational\,(1,3) or a bounded high-order polynomial, using the so-called 
data-driven method from~\cite{Zhou:2020cdt}, or using some theoretical frameworks and approaches, like DI$\chi$EFT. With such constrained 
functions, one could try to refit the entire $Q^{2}$ range of available data with an 
analytically well-behaved function~\cite{Kelly:2004hm}.

\section{Acknowledgements} 
This work is supported in part by the U.S. Department of Energy, Office of Science, Office of Nuclear Physics under Contract No. DE-FG02-03ER41231, 
as well as with Contract No. DE-AC05-06OR23177, under which the Jefferson Science Associates operates the Thomas Jefferson National Accelerator Facility.

\bibliography{Form-factor_Mainz_reanalysis_PRC}

\end{document}